\documentclass[pra,twocolumn,english,showpacs]{revtex4-1}
\usepackage{graphicx}
\usepackage{amssymb}
\usepackage{color}

\begin{document}

\title{Isotope shift and hyperfine splitting of the $4s \rightarrow 5p$ transition in potassium}

\author{Alexandra Behrle, Marco Koschorreck, and Michael K{\"o}hl}

\affiliation{Cavendish\,Laboratory,\,University~of Cambridge, JJ Thomson Avenue, Cambridge CB30HE, United Kingdom}

\begin{abstract}
We have investigated $4s\;^2S_{1/2} \rightarrow 5p\;^2P_{1/2}$ transition ($D_1$ line) of the potassium isotopes $^{39}$K, $^{40}$K, and $^{41}$K using Doppler-free laser saturation spectroscopy. Our measurements reveal the hyperfine splitting of the $5p\;^2P_{1/2}$ state of $^{40}$K and we have determined the specific mass shift and the nuclear field shift constants for the blue $D_1$ line.
\end{abstract}

\pacs{32.10.-f %Properties of atoms
31.30.Gs %Hyperfine interactions and isotope effects
}

\date{\today}

\maketitle

\section{Introduction}

The dependence of an atomic transition frequency on the properties of the nucleus is a frequently investigated question in laser spectroscopy. Possible effects include the hyperfine interaction, which relies on the coupling of the angular momenta of electrons and the nucleus, and isotope shifts, which depend on the mass of the nucleus and its internal charge distribution.

In atomic potassium, the isotopic composition makes systematic experiments of hyperfine interactions and isotope shifts of atomic transitions very challenging.
Two highly abundant isotopes ($^{39}$K and $^{41}$K) are accompanied by many isotopes with very low natural abundance. However, the knowledge of the relative frequency shifts of at least {\it two pairs} of isotopes are required to determine the size of the two non-trivial contributions to the isotope shift: the electronic correlation factor and the volume shift due to the charge distribution of the nucleus \cite{King1963}. Therefore, spectroscopy of the third-most abundant isotope $^{40}$K ($0.01\%$) is crucial. Aside from the prospect of exploring the fundamental atomic structure of potassium, demand for precise laser spectroscopic properties of $^{40}$K also stems from its use in the physics of degenerate Fermi gases,
where it is the isotope of choice in many experiments
\cite{Demarco1999c,Roati2002,Aubin2005,Kohl2005b,Ospelkaus2006,Rom2006,Klempt2007,Wille2008,Taglieber2008,Tiecke2010,Frohlich2011}.

The $4s \rightarrow 4p$ transition of potassium near 770\,nm has been investigated thoroughly several decades ago for isotopes
from $^{38}$K through $^{47}$K \cite{Bendali1981,Touchard1982}. The results of laser spectroscopic isotope shift measurements have been compared to nuclear
physics experiments regarding the nuclear charge radius as well as to theoretical calculations for the electronic correlations \cite{Martensson1990}. Meanwhile, frequency-comb based measurements of this transition have significantly improved the accuracy for the isotopes
$^{39}$K through $^{41}$K \cite{Falke2006}.

The {\it blue} $4s \rightarrow 5p$ transition of potassium near $\lambda=405$\,nm is considerably weaker than the $4s \rightarrow 4p$ transition, which
makes experiments on isotopes with small natural abundance even more difficult. Consequently, investigations of this transition are scarce.
However, recently interest in this transition has renewed because advances in semiconductor laser technology make the wavelength easily accessible \cite{Uetake2003}.
Measurements of the isotope shifts of the $4s \rightarrow 5p$ transition have been undertaken in $^{39}$K and $^{41}$K using Doppler-free saturation spectroscopy
\cite{Halloran2009} and in $^{39}$K, $^{40}$K, and $^{41}$K using resonance ionization mass spectrometry (RIMS) \cite{Iwata2010}.
While the sensitivity of the RIMS method is good to detect the low-abundant isotope $^{40}$K, its spectral resolution is limited.
In particular, the hyperfine splitting of the excited state $5s\;^2P_{1/2}$ was not observed for any of the isotopes.

\section{Experimental setup}

\begin{figure}[htbp]
\includegraphics[width=0.9\columnwidth,clip=true]{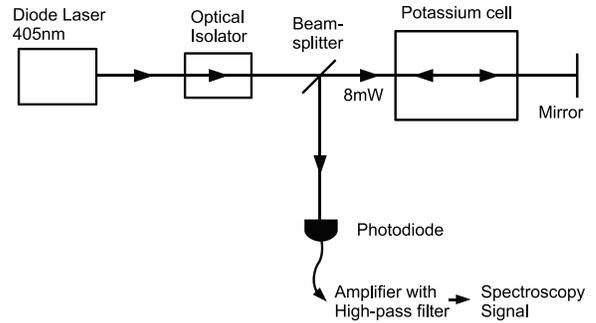}
  \caption{Schematic of the Doppler-free saturation spectroscopy setup.}
   \label{setup}
\end{figure}

We perform Doppler-free saturation spectroscopy on a potassium vapor cell containing isotopically enriched $^{40}$K ($4\%$). Our experimental setup is sketched in Figure \ref{setup}. The cell (length 100\,mm) is heated to approximately $120^\circ$\,C to provide sufficient absorption on the $4s\rightarrow 5p$ transition. The laser source is a diode laser (Mitsubishi ML320G2-11) operating at 405\,nm in an external cavity setup \cite{Ricci1995}. The laser frequency is swept with a frequency of 11\,Hz by angle-tuning the position of the diffraction grating in the laser setup using a piezo transducer.

The laser beam is sent through the vapor cell and is retro-reflected from a mirror. After traveling again through the cell, the beam is reflected by a beam splitter to a photodiode. The photodiode signal is amplified and high-pass filtered with a bandwidth of 1\,kHz. We use derivatives of Lorentzian functions to fit the measured data and to determine the zero crossing of each line. Afterwards we use the well known ground state splitting of $^{41}$K of 254.0\,MHz and $^{40}$K of 1285.8\,MHz \cite{Arimondo1977} as normalization of the frequency axis and for linearizing the piezo scan.

The power of the laser when entering the spectroscopy cell is approximately 8\,mW in a beam of 5\,mm diameter, therefore exceeding the saturation intensity $I_{sat}=(\pi h c)/(3 \lambda^3 \tau)\approx$ 3.4\,W/m$^2$ of this transition by two orders of magnitude. Here, $h$ is Planck's constant, $c$ is the speed of light, and $\tau= 930$\,ns is the lifetime of the excited state \cite{NIST2010}. For the $F=7/2 \rightarrow F'$ and $F=9/2 \rightarrow F'$ transitions of $^{40}$K we observe a line width of approximately 13\,MHz, somewhat larger than the expected saturation-broadened line width of 5\,MHz, which possibly could result from residual magnetic fields inside the unshielded vapor cell.

\begin{figure}[htbp]
\includegraphics[width=\columnwidth,clip=true]{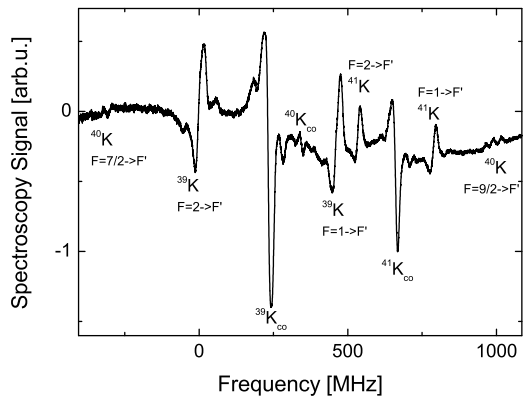}
  \caption{Doppler-free saturation spectrum of the isotopes $^{39}$K, $^{40}$K, and $^{41}$K on the $4S_{1/2} \rightarrow 5P_{1/2}$ transition. The zero of the frequency axis is chosen at the $F=2 \rightarrow F^\prime$ transition of $^{39}$K.}
 \label{spectrum}
\end{figure}

Figure \ref{spectrum} shows the measured spectrum of the three isotopes $^{39}$K, $^{40}$K, and $^{41}$K.
The zero of the frequency axis is chosen at the $F=2 \rightarrow F^\prime=1$ transition of $^{39}$K.
Here $F=I+J$ denotes the total angular momentum of the atom comprising of the nuclear spin $I$ ($I=3/2$ for $^{39}$K
and $^{41}$K, and $I=4$ for $^{40}$K) and electronic angular momentum $J=1/2$. We use the notation $F$ to label the $4s\;^2S_{1/2}$
ground state and $F^\prime$ to label the $5p\;^2P_{1/2}$ excited state. We estimate the uncertainty of the frequency calibration to be $0.6$\,MHz.

\section{Hyperfine splitting of the $5p\;^2P_{1/2}$ state of $^{40}$K}

\begin{figure}[htbp]
\includegraphics[width=\columnwidth,clip=true]{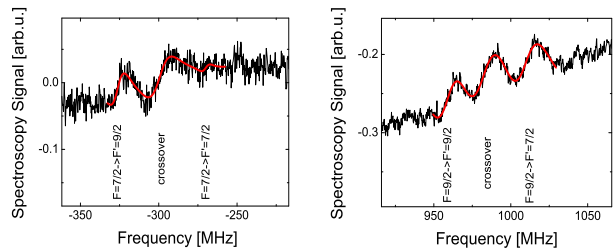}
  \caption{(Color online) Close-up view of the Doppler-free signal of $^{40}$K revealing the hyperfine splitting.
  The solid lines show the fit to the data to extract the positions of the lines.}
 \label{hfs}
\end{figure}

While for the two most abundant isotopes $^{39}$K and $^{41}$K the hyperfine splitting of the $5p\;^2P_{1/2}$ state has been measured previously \cite{Arimondo1977},
no such data have been available for $^{40}$K. Our Doppler-free saturation spectrum directly reveals the hyperfine splitting of the $5p\;^2P_{1/2}$ state of $^{40}$K,
see Figure \ref{spectrum}, on transitions originating from both the ground states $F=7/2$ and $F=9/2$. Figure \ref{hfs} shows a close-up of the two sections together with our fits. On the $F=9/2 \rightarrow F'$ manifold we observe three equally strong lines, which correspond to the $F=9/2 \rightarrow F'=9/2$ and $F=9/2 \rightarrow F'=7/2$ transitions together with a crossover resonance in between. On the $F=7/2 \rightarrow F'$ manifold we observe two strong lines, which we assign to the $F=7/2 \rightarrow F'=9/2$ transition and a crossover resonance,
and a very weak signal which we assign to the $F=7/2 \rightarrow F'=7/2$ transition. Theoretically, this line has a line strength which is factor of five smaller than the $F=7/2 \rightarrow F'=9/2$ line which is compatible with our data. From the splitting of the lines we determine the hyperfine interaction constant $A_{40}^{5p\;^2P_{1/2}}=(-12.0 \pm 0.9)$\,MHz.
Our measured value is in good agreement with the value obtained by scaling the previously known value $A_{39}^{5p\;^2P_{1/2}}=(9.02 \pm 0.17)$\,MHz \cite{Arimondo1977}
by the known ratio of the nuclear g-factors $g_{I,40}/g_{I,39}=-1.24$ which would result in $A_{40}^{5p\;^2P_{1/2}}=(-11.2\pm 0.2)$\,MHz. The much smaller excited state hyperfine splitting of the isotopes $^{39}$K and $^{41}$K is below our resolution.

\section{Isotope shifts}

The isotope shift $\delta \nu^\alpha_{i,j}$ of an atomic transition $\alpha$ between two different isotopes labeled $i$ and $j$ arises from a sum of contributions
of different physical origins, which can be parameterized by the following expression \cite{Martensson1990}

\begin{equation}
\delta \nu^\alpha_{i,j}=  (k^\alpha_{NMS}+ k^\alpha_{SMS}) \frac{(M_j-M_i)}{M_i(M_j+m_e)} + F^\alpha \delta\langle r^2 \rangle_{i,j}.
\label{eqn1}
\end{equation}

Here, $m_e$ is the electron mass, $M_i$ the mass of the lighter nucleus and $M_j$ the mass of the heavier nucleus. This choice defines the convention of the sign of the shift. The first contribution is the normal mass shift (NMS) which results from the change of the reduced mass in the effective two-body problem due to the change of the mass of the nucleus. $k_{NMS}^\alpha= \nu_i^\alpha m_e$, with $\nu_i^\alpha=740.529$\,THz \cite{NIST2010} denoting the resonance frequency of the transition. Usually, the normal mass shift is the dominant contribution to the observed isotope shift.

The remaining shift, often referred to as the residual mass shift $\delta \nu^\alpha_{RMS,(i,j)}=k^\alpha_{SMS} \frac{(M_j-M_i)}{M_i(M_j+m_e)} + F^\alpha \delta\langle r^2 \rangle_{i,j}$, results from more subtle effects. The constant $k^\alpha_{SMS}$ describing the specific mass shift is the expectation value of the operator $\sum_{i>j} \textbf{p}_i \textbf{p}_j$ of the different electronic momenta $\textbf{p}_i$. Generally, it is difficult to evaluate this contribution theoretically for atoms with many electrons. However, since $k^\alpha_{SMS}$ depends only on electron-electron correlations, it is a constant for a chosen transition $\alpha$ and does not depend on the nuclear mass. In contrast, the contribution of the field shift $F^\alpha \delta\langle r^2 \rangle_{i,j}$ is determined by the difference of the electron density at the nucleus for two involved states times the difference of the charge radius of the nucleus for the two isotopes. For determining both constants $k^\alpha_{SMS}$ and $F^\alpha$ at least two pairs of isotopes have to be measured if the nuclear charge radii are known.

Our spectrum (see Figure \ref{spectrum}) contains all three isotopes $^{39}$K, $^{40}$K, and $^{41}$K and we determine the isotope shifts. We find $\Delta \nu^{4s\rightarrow 5p}_{39,41}=(454.2\pm 0.8)$\,MHz, which is in agreement with the results of Halloran et al. \cite{Halloran2009} and Iwata et al. \cite{Iwata2010}. The isotope shift of the low-abundance isotope $^{40}$K with respect to $^{39}$K is $\Delta \nu^{4s \rightarrow 5p}_{39,40}=(235.0\pm 2.0)$\,MHz. Our value for this isotope shift is in disagreement with the very recent result of $(207\pm13)$\,MHz \cite{Iwata2010}. This discrepancy could be due to the fact that the hyperfine structure was not resolved in experiment \cite{Iwata2010} or because of additional systematic errors, which were estimated up to another $10\%$ in their setup. From the measured isotope shift we determine the residual mass shift $\Delta \nu^\alpha_{RMS,(i,j)}$. For the residual mass shift we find $\Delta\nu^{4s \rightarrow 5p}_{RMS,(39,41)}=(-54.4\pm 0.8)$\,MHz and $\Delta \nu^{4s\rightarrow 5p}_{RMS,(39,40)}=(-26.0\pm 2.0)$\,MHz.

Using the nuclear charge radii differences for potassium tabulated in reference \cite{Martensson1990} we determine $k^{4s\rightarrow 5p}_{SMS}=(-39\pm5)$\,GHz amu and $F^{4s\rightarrow 5p}=(-43\pm55)$\,MHz fm$^{-2}$ for the $4s\rightarrow 5p$ transition. The error bars are largely determined by the uncertainty of the $\Delta \nu^{4s\rightarrow 5p}_{RMS,(39,40)}$ frequency shift. Our results should be compared to $k^{4s\rightarrow 4p}_{SMS}=(-15.4\pm 3.8)$\, GHz amu and $F^{4s\rightarrow4p} = (-110\pm 3)$\,MHz fm$^{-2}$ on the $4s\rightarrow 4p$ transition \cite{Martensson1990}. We conclude that for the $4s \rightarrow 5p$ transition the specific isotope shift $k_{SMS}$ is larger, possibly because the higher electronically excited state gives rise to larger electron-electron correlations. In contrast, the field shift reduces as compared to the $4s \rightarrow 4p$ transition which could reflect the smaller overlap of the $5p$ wave function with the nucleus as compared to the $4p$ electronic wave function.

\section{Conclusion}

In conclusion, we have measured the hyperfine splitting of the $5p\;^2P_{1/2}$ state of $^{40}$K and determined the isotope shift of the $4s\;^2S_{1/2} \rightarrow 5p\;^2P_{1/2}$ transition, exceeding the previously achieved accuracy for the low-abundance isotope $^{40}$K by one order of magnitude. Our results contribute to determining the atomic structure of potassium more accurately. From the measurement of two isotope shifts we could determine the specific mass shift constant $k^{4s\rightarrow 5p}_{SMS}=(-39\pm5)$\,GHz amu and field shift constant $F^{4s\rightarrow 5p}=(-43\pm55)$\,MHz fm$^{-2}$ of the $4s\rightarrow 5p$ transition. The hyperfine interaction constant of $(-12\pm 0.9)$\,MHz is in agreement with the theoretically anticipated value.

The work has been supported by {EPSRC} (EP/G029547/1) and {ERC} (Grant number 240335).

%\bibliography{thesis}

%

\end{document}